\documentclass[12pt, reqno, letter]{amsart}
\usepackage{setspace}
\usepackage[foot]{amsaddr}

\usepackage{pkgfile}

\allowdisplaybreaks

\title[Effect Modification in Observational Studies]{A Powerful Approach to the Study of Moderate Effect Modification in Observational Studies}

\author{Kwonsang Lee$^{1}$, Dylan S. Small$^{2}$, and Paul R. Rosenbaum$^{2}$}
\address{$^{1}$Department of Biostatistics, Harvard School of Public Health}
\address{$^{2}$Department of Statistics, University of Pennsylvania}

\begin{document}

\begin{abstract} 
Effect modification means the magnitude or stability of a treatment effect varies as a function of an observed covariate. Generally, larger and more stable treatment effects are insensitive to larger biases from unmeasured covariates, so a causal conclusion may be considerably firmer if this pattern is noted if it occurs. We propose a new strategy, called the submax-method, that combines exploratory and confirmatory efforts to determine whether there is stronger evidence of causality --- that is, greater insensitivity to unmeasured confounding --- in some subgroups of individuals. It uses the joint distribution of test statistics that split the data in various ways based on certain observed covariates. For $L$ binary covariates, the method splits the population $L$ times into two subpopulations, perhaps first men and women, perhaps then smokers and nonsmokers, computing a test statistic from each subpopulation, and appends the test statistic for the whole population, making $2L+1$ test statistics in total. Although $L$ binary covariates define $2^{L}$ interaction groups, only $2L+1$ tests are performed, and at least $L+1$ of these tests use at least half of the data.  The submax-method achieves the highest design sensitivity and the highest Bahadur efficiency of its component tests. Moreover, the form of the test is sufficiently tractable that its large sample power may be studied analytically. The simulation suggests that the submax method exhibits superior performance, in comparison with an approach using CART, when there is effect modification of moderate size. Using data from the NHANES I Epidemiologic Follow-Up Survey, an observational study of the effects of physical activity on survival is used to illustrate the method.  The method is implemented in the $\texttt{R}$ package $\texttt{submax}$ which contains the NHANES example.  An on-line appendix provides simulation results and further analysis of the example.
\end{abstract}

\keywords{Causal effects; causal inference; design sensitivity;	effect modification; epidemiology; observational study; sensitivity analysis; testing twice.}

\doublespacing

\maketitle

\section{Does Physical Activity Prolong Life? Equally for Everyone?}

\subsection{A Matched Comparison of Physical Inactivity and Survival}
\label{ssIntroEG}

Davis et al. (1994) used the NHANES I Epidemiologic follow-up study (NHEFS) to ask: Is greater physical activity reported at the time of the NHANES I survey associated with a longer subsequent life? We examine the same data in a similar way, but with new methodology, specifically the subgroup maximum method or submax-method.

The NHANES I sample was interviewed in 1971-1975 and followed for survival until 1992. Physical activity was measured in two variables: self-reported nonrecreational activity and self-reported recreational activity. We formed a treated group of 470 adults who were ``quite inactive'', both at work and at leisure, and we matched them to a control group of 470 adults who were quite active (``very active'' in physical activity outside of recreation and ``much'' or ``moderate'' recreational activity). We compare quite inactive to quite active because making the treated and control groups sharply differ in dose increases the insensitivity of the study to unobserved confounding.  More precisely, if, in a large study, there was no unmeasured bias together with a treatment effect exhibiting larger effects at higher doses, then a study of high dose versus no dose would report greater insensitivity to unmeasured bias (Rosenbaum, 2004).  Following Davis et al. (1994), we excluded people who were quite ill at the time of the NHANES I survey. We included people aged between 45 and 74 at baseline, and excluded people who, prior to NHANES I, had heart failure, a heart attack, stroke, diabetes, polio or paralysis, a malignant tumor, or a fracture of the hip or spine.

\begin{table}
\caption{\label{tabCov} Covariate balance in 470 matched, treatment-control pairs. The standardized difference (Std. Dif) is the difference in means before and after matching in units of the standard deviation before matching.
The 470 controls After matching were selected from 1482 potential controls Before matching.  The matching was exact for sex, poverty, and current smoking, and controlled other covariates by minimizing the
total Mahalanobis distance within matched pairs.  The informal, unpaired $P$-values compare covariate balance attained by matching to the balance anticipated in a completely randomized experiment.
}
\centering
\begin{tabular}{l|cc|c|cc}
\hline
& \multicolumn{2}{c|}{Covariate Mean} & & \multicolumn{2}{c}{Std. Dif.}\\
Covariate & Treated & Control & $P$-value & Before & After \\ \hline
Age & 61.7 & 61.7 & 0.985 & 0.283 & 0.001 \\
Male & 0.415 & 0.415 & 1.000 & -0.245 & 0.000 \\
White & 0.789 & 0.823 & 0.187 & -0.252 & -0.093 \\ \hline
Poverty & 0.460 & 0.460 & 1.000 &  0.377 & 0.000\\
Former Smoker & 0.170 & 0.145 & 0.283 & -0.142 & 0.064 \\
Current Smoker & 0.360 & 0.360 & 1.000 & -0.141 & 0.000\\
Working last three months & 0.247 & 0.247 & 1.000 & -0.589 & 0.000\\
Married & 0.621 & 0.666 & 0.153 & -0.350 & -0.099\\
Dietary Adequacy & 3.254 & 3.379 & 0.143 &  -0.303 & -0.098\\ \hline
& \multicolumn{5}{|c}{Education} \\ \hline$\>$ $\le
$ 8 & 0.494 & 0.466 & 0.397 & 0.309 & 0.057 \\
$\>$ 9-11 & 0.183 & 0.204 & 0.410 & -0.097 & -0.053\\
$\>$ High School & 0.166 & 0.172 & 0.794 & -0.193 & -0.016 \\
$\>$ Some College & 0.066 & 0.070 & 0.796 & -0.158 & -0.015\\
$\>$ College & 0.085 & 0.085 & 1.000 &  0.038 & 0.000\\
$\>$ Missing & 0.006 & 0.002 & 0.317 & 0.004 & 0.054 \\
\hline
& \multicolumn{5}{|c}{Alcohol Consumption} \\ \hline
$\>$ Never & 0.406 & 0.432 & 0.428 & 0.189 & -0.053 \\
$\>$ $<1$ time per month & 0.198 & 0.185 & 0.619 & 0.016 & 0.032  \\
$\>$ 1-4 times per month & 0.172 & 0.153 & 0.427 & -0.125 & 0.048\\
$\>$ 2+ times per week & 0.089 & 0.089 & 1.000 & -0.069 & 0.000 \\
$\>$ Just about everyday/everyday & 0.134 & 0.140 & 0.776 & -0.073 & 0.000 \\
\hline
\end{tabular}
\end{table}

Table~\ref{tabCov} shows the matched covariates. Pairs were exactly matched on sex, smoking status (current smoker) and income (cut at $2\times$ the poverty level). Other matched variables were age, race (white or other), years of education, employed or not during the previous three months, marital status, alcohol consumption and dietary quality (number of five nutrients -- protein, calcium, iron, Vitamin A and Vitamin C -- that were consumed at more than two thirds of the recommended dietary allowance). After matching, the groups are similar. Before matching, the inactive group was older, more often female, more often nonwhite, more often poor, more often not working, more often not married, and less often had an adequate diet.

\begin{figure}
\centering
\includegraphics[width=140mm]{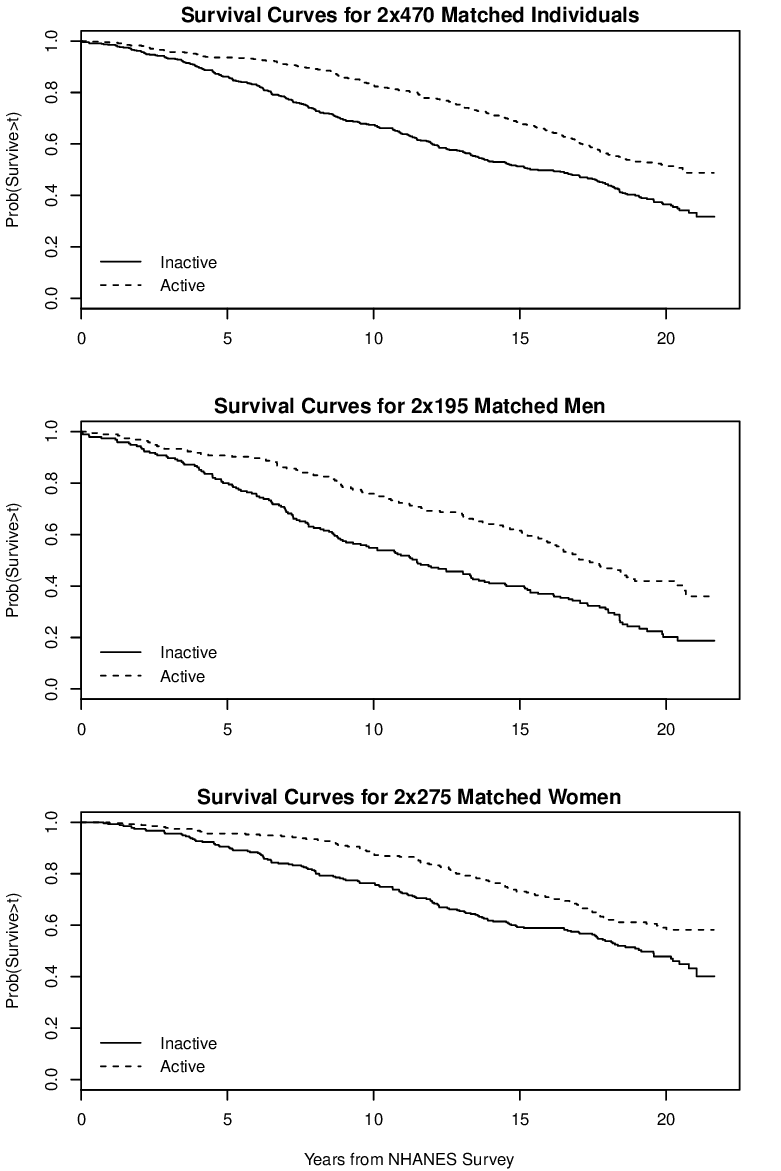}
\caption{Kaplan-Meier survival curves for inactive individuals and their matched active controls, for all 470 pairs, for 195 pairs of men and for 275 pairs of women.}
\end{figure}

The top of Figure 1 shows survival in matched active and inactive groups. We ask: (i) What magnitude of unmeasured bias from nonrandom treatment assignment would need to be present to explain Figure 1 as something other than an effect caused by inactivity? (ii) Is there greater insensitivity to unmeasured bias in some subgroups because the ostensible effect is larger in those subgroups, or is there similar evidence of effect in all subgroups? We study sex, smoking and income as potential effect modifiers.

Wager and Athey (2017) use random forests to estimate average treatment effects as they vary with covariates.  Zhao et al. (2017) draw inferences about the average treatment effect for covariates selected using the data.  Neither article considers sensitivity to unmeasured confounding, which is a main focus here.  Effect modification has consequences for sensitivity to unmeasured biases, a central concern in observational studies.

\subsection{A New Approach to Effect Modification in Observational Studies}

If some subgroups experience larger or more stable effects, then the ostensible effect of a treatment may be less sensitive to bias from nonrandomized treatment assignment in these subgroups; see Hsu et al. (2013). Conversely, if a treatment appears to be highly effective in all subgroups, then it is safer to generalize to other populations that may have different proportions of people in the various subgroups.

One approach to effect modification constructs a few promising subgroups from several measured covariates using, say, the CART technique of Breiman, Friedman, Olshen and Stone, as discussed by Hsu et al. (2013, 2015), and as described in on-line \S 3.6. A limitation of this approach is that it is hard to study the power of such a technique except by simulation, because the CART step does not lend itself to such an evaluation. In the current paper we propose a different approach --- the submax method --- for which a theoretical evaluation is possible. The submax method has formulas for power and design sensitivity, and permits statements about Bahadur efficiency. The new submax method achieves the largest --- i.e., best --- of the design sensitivities for the subgroups, and the highest Bahadur efficiency of the subgroups; moreover, both the power formula and a simulation confirm that the asymptotic results are a reasonable guide to performance in samples of practical size. The simulation in on-line \S 3.6 also compares the submax and CART methods. An additional limitation of the CART method is that it is only defined for matched pairs, not for matched sets. In contrast, the submax-method works for pairs, for matched sets with multiple controls, variable numbers of controls and with ``full matching'' as in Hansen and Klopfer (2012).

The submax-method considers a single combined analysis together with several ways to split the population into subgroups. It does not form the interaction of subgroups, which would quickly become thinly populated with small sample sizes; rather, it considers one split, reassembles the population, then considers another split. If the splits were defined by $L$ binary covariates, then there would be $2^{L}$ interaction subgroups, but the submax-method would do only 1 overall test plus $2L$ subgroup tests, making a total of $2L+1$ highly correlated tests, not $2^{L}$ independent tests. If the binary covariates each split every subpopulation in half, then each interaction subgroup would contain a fraction $2^{-L}$ of the population --- i.e., not much --- but each of our $2L$ subgroup tests would use half the population --- i.e., a much larger fraction. The submax-method uses the joint distribution of the $2L+1$ test statistics, with the consequence that the correction for multiple testing is quite small due to the high correlation among the test statistics. Specifically, the two halves of one binary split are independent because they refer to different people, but each of those test statistics is highly correlated with test statistics for other splits, because all the splits use the same people. In the example, we split the population by gender (male or female), by current cigarette smoking (yes or no), and by two income groups, so we do $2L+1=2\times3+1=7$ correlated tests. The statistics for men and women are independent, but the statistics for men and smokers are highly correlated because there are many male smokers.

\section{Notation and Review of Observational Studies}
\label{secNotation}

\subsection{Treatment Effects in Randomized Experiments}
\label{ssNotationCausal}

There are $G$ groups, $g=1,\ldots,G$, of matched sets, $i=1$, \ldots, $I_{g}$, with $n_{gi}$ individuals in set $i$, $j=1,$ $\ldots$, $n_{gi}$, one treated individual with $Z_{gij}=1$ and $n_{gi}-1$ controls with $Z_{gij}=0$, so that $1=\sum_{j=1}^{n_{gi}}Z_{gij}$ for each $g$, $i$. Write $I_{+}=\sum_{g=1}^{G}I_{g}$.  By design, matched sets are disjoint: no individual appears in more than one matched set. Matched sets were formed by matching for an observed covariate $x_{gij}$, but may fail to control an unobserved covariate $u_{gij}$, so that $x_{gij}=x_{gik}$ for each $g$, $i$, $j$, $k$, but possibly $u_{gij}\neq u_{gik}$. In \S \ref{ssIntroEG}, the matched sets are pairs, $n_{gi}=2$, and there are $G=2^{3}=8$ groups of pairs defined by combinations of $L=3$ binary covariates, sex, smoking and income group, with $I_{+}=470$ pairs in total.

In the Neyman-Rubin notation, individual $gij$ exhibits response $r_{Tgij}$ if treated or response $r_{Cgij}$ under control, so she exhibits response $R_{gij}=Z_{gij}\,r_{Tgij}+\left(  1-Z_{gij}\right) \, r_{Cgij}$, and the causal effect, $r_{Tgij}-r_{Cgij}$, is not observed. Fisher's hypothesis of no effect asserts that $H_{0}:r_{Tgij}=r_{Cgij}$ for all $i$, $j$.  Fisher's randomization test of $H_{0}$ is the same as the permutation test of the hypothesis of equal distributions of responses within matched sets; see Lehmann and Romano (2005, \S5.8).  Write $\mathcal{F}=\left\{  \left(  r_{Tgij},r_{Cgij},x_{gij},u_{gij} \right), \, g=1,\ldots,G,\,i=1,\ldots,I_{g},\,j=1,\ldots,n_{gi}\right\}$. Write $\left\vert \mathcal{S}\right\vert $ for the number of elements in a finite set $\mathcal{S}$.

Write $\mathcal{Z}$ for the set containing the $\left\vert \mathcal{Z}\right\vert = \prod_{g=1}^{G} \prod_{i=1}^{I_{g}} n_{gi}$ possible values $\mathbf{z}$ of the treatment assignment $\mathbf{Z}=\left(  Z_{111},Z_{112},\ldots,Z_{G,I_{G},n_{G,I_{G}}}\right)^{T}$, so $\mathbf{z}\in\mathcal{Z}$ if $z_{gij}=0$ or $z_{gij}=1$ and $1=\sum_{j=1}^{n_{gi}}z_{gij}$ for each $gi$. Conditioning on the event $\mathbf{Z}\in\mathcal{Z}$ is abbreviated as conditioning on $\mathcal{Z}$. In an experiment, randomization picks a $\mathbf{Z}$ at random from $\mathcal{Z}$, so that $\Pr\left(  \left.  \mathbf{Z}=\mathbf{z\,}\right\vert \,\mathcal{F},\,\mathcal{Z}\right)  =\left\vert \mathcal{Z}\right\vert ^{-1}$ for each $\mathbf{z}\in\mathcal{Z}$. In a randomized experiment, randomization creates the exact null randomization distribution of familiar test statistics, such as Wilcoxon's signed rank statistic or the mean pair difference or Maritz (1979)'s version of Huber M-statistic. In the analysis of the paired censored survival data in \S \ref{ssIntroEG}, the test statistic is the Prentice-Wilcoxon test of O'Brien and Fleming (1987). These test statistics and many others are of the form $T=\sum_{g=1}^{G}\sum_{i=1}^{I_{g}}\sum_{j=1}^{n_{gi}}Z_{gij}\,q_{gij}$ for suitable scores $q_{gij}$ that are a function of the $R_{gij}$, $n_{gi}$ and possibly the $x_{gij}$, so that, under $H_{0}$ in a randomized experiment, the conditional distribution $\Pr\left(  \left.  T\mathbf{\,}\right\vert \,\mathcal{F},\,\mathcal{Z}\right)  $ of the test statistic $T$ is the distribution of the sum of fixed scores $q_{gij}$ with $Z_{gij}=1$ selected at random.

In large sample approximations, the number of groups, $G$, will remain fixed, and the number of matched sets $I_{g}$ in each group will increase without bound.

\subsection{Sensitivity to Unmeasured Biases in Observational Studies}
\label{ssNotationSensitivity}

In an observational study, conventional tests of $H_{0}$ appropriate in the randomized experiments in \S \ref{ssNotationCausal} can falsely reject a true null hypothesis of no effect because treatments are not assigned at random, $\Pr\left(  \left.  \mathbf{Z}=\mathbf{z\,}\right\vert \,\mathcal{F},\,\mathcal{Z}\right)  \neq\left\vert \mathcal{Z}\right\vert ^{-1}$. A simple model for sensitivity analysis in observational studies assumes that, in the population prior to matching for $x$, treatment assignments are independent and two individuals, $gij$ and $g^{\prime}i^{\prime}j^{\prime}$, with the same observed covariates, $x_{gij}=x_{g^{\prime}i^{\prime}j^{\prime}}$, may differ in their odds of treatment by at most a factor of $\Gamma\geq 1$,
\begin{equation}
\frac{1}{\Gamma}\leq\frac{\Pr\left(  \left.  Z_{gij}=1\mathbf{\,}\right\vert\,\mathcal{F}\right) \,\Pr\left(  \left.  Z_{g^{\prime}i^{\prime}j^{\prime}}=0\mathbf{\,}\right\vert \,\mathcal{F}\right)  }{\Pr\left(  \left. Z_{g^{\prime}i^{\prime}j^{\prime}}=1\mathbf{\,}\right\vert \,\mathcal{F}\right)  \,\Pr\left(  \left.  Z_{gij}=0\mathbf{\,}\right\vert \,\mathcal{F}\right)  }\leq\Gamma\text{ whenever }x_{gij}=x_{g^{\prime}i^{\prime}j^{\prime}}\text{;}
\label{eqSenMod}
\end{equation}
then the distribution of $\mathbf{Z}$ is returned to $\mathcal{Z}$ by conditioning on $\mathbf{Z}\in\mathcal{Z}$.

Under the model (\ref{eqSenMod}), one obtains conventional randomization inferences for $\Gamma=1$, but these are replaced by an interval of $P$-values or an interval of point estimates or an interval of endpoints for a confidence interval for $\Gamma>1$. The intervals become longer as $\Gamma$ increases, the interval of $P$-values tending to $\left[  0,1\right]  $ as $\Gamma \rightarrow\infty$, reflecting the familiar fact that association, no matter how strong, does not logically entail causation. At some point, the interval is sufficiently long to be uninformative, for instance including $P$-values that would both reject and accept the null hypothesis of no effect. The question answered by a sensitivity analysis is: How much bias in treatment assignment, measured by $\Gamma$, would need to be present before the study becomes uninformative? For instance, how large would $\Gamma$ have to be to produce a $P$-value above $\alpha$, conventionally $\alpha=0.05$?

An approximation to the upper bound on the $P$-value is obtained as follows; see Gastwirth, Krieger and Rosenbaum (2000) for detailed discussion and see Rosenbaum (2007, \S 4; 2014) for its application to Huber-Maritz M-tests. Assume $H_{0}$ is true for the purpose of testing it, so that $R_{gij}=r_{Cgij}$ and $q_{gij}$ are fixed by conditioning on $\mathcal{F}$. Write $T_{g}=\sum_{i=1}^{I_{g}}\sum_{j=1}^{n_{gi}} Z_{gij}\,q_{gij}$, so that $T=\sum_{g=1}^{G}T_{g}$. Subject to (\ref{eqSenMod}) for a given $\Gamma\geq1$, find the maximum expectation, $\mu_{\Gamma g}$, of $T_{g}$. Also, among all treatment assignment probabilities that satisfy (\ref{eqSenMod}) and that achieve the maximum expectation $\mu_{\Gamma g}$, find the maximum variance, $\nu_{\Gamma g}$, of $T_{g}$. If $T\geq$ $\sum_{g=1}^{G}\mu_{\Gamma g}$, report as the upper bound on the $P$-value for $T$,
\begin{equation}
1-\Phi\left\{  \left(  \sum_{g=1}^{G}T_{g}-\mu_{\Gamma g}\right)  / \sqrt{\sum_{g=1}^{G}\nu_{\Gamma g}}\right\}  \text{,}
\label{eqSenBound}
\end{equation}
where $\Phi\left(  \cdot\right)  $ is the standard Normal cumulative distribution. The bound is derived as $\min\left(  I_{g}\right) \rightarrow\infty$ with some mild conditions to ensure that no one $q_{gij}$ dominates the rest, and that the fixed scores $q_{gij}$ do not become degenerate as $\min\left(  I_{g}\right)  $ increases. For $\Gamma=1$, this yields a Normal approximation to a randomization $P$-value using $T$ as the test statistic. If treatment assignments were governed by the probabilities satisfying (\ref{eqSenMod}) that yield $\mu_{\Gamma g}$ and $\nu_{\Gamma g}$, then, under $H_{0}$ and mild conditions on the $q_{gij}$, the joint distribution of the $G$ statistics $\left(  T_{g}-\mu_{\Gamma g} \right) / \nu_{\Gamma g}^{1/2} $,
converges to a $G$-dimensional Normal distribution with expectation $\mathbf{0}$ and covariance matrix $\mathbf{I}$ as $\min\left(I_{g}\right)  \rightarrow\infty$. Simpler methods of proof and formulas apply in simple cases, such as matched pairs; for instance, contrast \S 3 and \S 4 of Rosenbaum (2007). Write $\boldsymbol{\mu}_{\Gamma}=\left(  \mu_{\Gamma1},\ldots\mu_{\Gamma G}\right)^{T}$ and $\mathbf{V}_{\Gamma}$ for the $G\times G$ diagonal matrix with $g$th diagonal element $\nu_{\Gamma g}$.

For various methods of sensitivity analysis, see Egleston et al. (2009), Gilbert et al. (2003), Hosman et al. (2010), and Liu et al. (2013).

\subsection{Design Sensitivity and Bahadur Efficiency}

Suppose there is a treatment effect with no bias from $u_{gij}$, and call this the favorable situation. If an investigator were in the favorable situation, then she would not know it, and the best she could hope to say is that the results are insensitive to moderate biases $\Gamma$. The power of a sensitivity analysis is the probability that she will be able to say this. In the favorable situation, the power of a level $\alpha$ sensitivity analysis at sensitivity parameter $\Gamma$ is the probability that (\ref{eqSenBound}) will be less than or equal to $\alpha$ when computed at the given $\Gamma$.

As $I_{+} \rightarrow\infty$, there is a value, $\widetilde{\Gamma}$, called the design sensitivity, such that the power tends to 1 if $\Gamma<\widetilde{\Gamma}$ and the power tends to zero if $\Gamma>\widetilde{\Gamma}$, so $\widetilde{\Gamma}$ is the limiting sensitivity to unmeasured bias for a given favorable situation and test statistic; see Rosenbaum (2004; 2010, Part III), Zubizarreta et al. (2013) and Stuart and Hanna (2013). In a favorable situation, for a specific $\Gamma$, the rate at which (\ref{eqSenBound}) declines to zero as $I_{+} \rightarrow\infty$ yields the Bahadur efficiency of the sensitivity analysis, which drops to zero at $\Gamma=\widetilde{\Gamma}$; see Rosenbaum (2015).

\section{Joint Bounds for Two or More Comparisons}
\label{secSubmax}

\subsection{Subgroup Comparisons}
\label{ssSubgroupComparisons}

There are $K$ specified comparisons, $k=1,\ldots,K$, involving $G$ groups of matched sets. A single comparison is a fixed nonzero vector $\mathbf{c}_{k}=\left(  c_{1k},\ldots,c_{Gk}\right)  ^{T}$ of dimension $G$ with $c_{gk}\geq0$ for $g=1,\ldots,G$, and we evaluate a comparison using the statistic $S_{k}=$ $\sum_{g=1}^{G}c_{gk}\,T_{g}$. The comparison $\mathbf{c}_{1}=\left(  1,\ldots,1\right)  ^{T}$ yields the overall test in \S \ref{ssNotationSensitivity}. By replacing the scores $q_{gij}$ in \S \ref{ssNotationSensitivity} by scores $q_{gij}^{\ast}=c_{gk}\,q_{gij}$, the bound for $S_{k}$ is obtained in parallel with (\ref{eqSenBound}). If groups $1$, \ldots, $G/2$ are matched sets of men and groups $G/2+1$, \ldots, $G$ are sets of women, then comparison $\mathbf{c}_{2}=\left(1,\ldots,1,0,\ldots,0\right)  ^{T}$ confines attention to men, while comparison $\mathbf{c}_{3}=\left(  0,\ldots,0,1,\ldots,1\right)^{T}$ confines attention to women.  In brief, we test the hypothesis of no treatment effect at all, plus the $2L$ hypotheses of no effect in $2L$ overlapping subpopulations.

If the treatment effect for women were larger than for men, the comparison, $\mathbf{c}_{3}$, restricted to women might be insensitive to larger unmeasured biases than the overall comparison, $\mathbf{c}_{1}$. In Hsu et al. (2013), a treatment to prevent malaria is far more effective for children than for adults, so that only very large biases could explain the ostensible benefits for children.

We describe a one-sided testing procedure, testing no effect against a positive effect at level $\alpha$.  A level $\alpha$ two-sided test uses the procedure twice at level $\alpha/2$, rejecting a component null hypothesis of no effect if it is rejected in either the positive or the negative direction.  In principle, there could be a positive treatment effect for men, a negative effect for women, where neither is apparent when everyone is pooled in a single test.

\subsection{Joint Evaluation of Subgroup Comparisons}
\label{ssDefineDmax}

Let $\mathbf{C}$ be the $K\times G$ matrix whose $K$ rows are the $\mathbf{c}_{k}^{T}=\left(  c_{1k},\ldots,c_{Gk}\right)  $, $k=1,\ldots,K$. Define $\boldsymbol{\theta}_{\Gamma}=\mathbf{C}\boldsymbol{\mu}_{\Gamma}$ and $\boldsymbol{\Sigma}_{\Gamma}=\mathbf{CV}_{\Gamma}\mathbf{C}^{T}$, noting that $\boldsymbol{\Sigma}_{\Gamma}$ is not typically diagonal. Write $\theta_{\Gamma k}$ for the $k$th coordinate of $\boldsymbol{\theta}_{\Gamma}$ and $\sigma_{\Gamma k}^{2}$ for the $k$th diagonal element of $\boldsymbol{\Sigma}_{\Gamma}$. Define $D_{\Gamma k}=\left(  S_{k}-\theta_{\Gamma k}\right)  /\sigma_{\Gamma k}$ and $\mathbf{D}_{\Gamma}=\left(  D_{\Gamma1},\ldots,D_{\Gamma K}\right)  ^{T}$. Finally, write $\boldsymbol{\rho}_{\Gamma}$ for the $K\times K$ correlation matrix formed by dividing the element of $\boldsymbol{\Sigma}_{\Gamma}$ in row $k$ and column $k^{\prime}$ by $\sigma_{\Gamma k}\,\sigma_{\Gamma k^{\prime}}$. Subject to (\ref{eqSenMod}) under $H_{0}$, at the treatment assignment probabilities that yield the $\mu_{\Gamma g}$ and $\nu_{\Gamma g}$, the distribution of $\mathbf{D}_{\Gamma}$ is converging to a Normal distribution, $N_{K}\left( \mathbf{0},\boldsymbol{\rho}_{\Gamma}\right)  $, with expectation $\mathbf{0}$ and covariance matrix $\boldsymbol{\rho}_{\Gamma}$ as $\min\left( I_{g}\right)  \rightarrow\infty$. Using this null distribution, the null hypothesis $H_{0}$ is tested using $D_{\Gamma\max}=\max_{1\leq k\leq K}D_{\Gamma k}$.  The $\alpha$ critical value $\kappa_{\Gamma,\alpha}$ for $D_{\Gamma\max}$ solves \begin{equation}
1-\alpha=\Pr\left(  D_{\Gamma\max}<\kappa_{\Gamma,\alpha}\right)  =\Pr\left( \frac{S_{k}-\theta_{\Gamma k}}{\sigma_{\Gamma k}}<\kappa_{\Gamma,\alpha},\,k=1,\ldots,K\right)  \label{eqCritConst}
\end{equation}
under $H_{0}$. In general, $\kappa_{\Gamma,\alpha}$ depends upon both $\Gamma$ and $\alpha$.  The multivariate Normal approximation to $\kappa_{\Gamma,\alpha}$ is obtained using the \texttt{qmvnorm} function in the \texttt{mvtnorm} package in R, as applied to the $N_{K}\left(  \mathbf{0},\boldsymbol{\rho}_{\Gamma}\right)$ distribution; see Genz and Bretz (2009). Notice that this approximation to $\kappa_{\Gamma,\alpha}$ depends upon $\Gamma$ only through $\boldsymbol{\rho}_{\Gamma}$, which in turn depends upon $\Gamma$ only through $\nu_{\Gamma g}$. The critical value $\kappa_{\Gamma,\alpha}$ for $D_{\Gamma\max}$ is larger than $\Phi^{-1}\left(  1-\alpha\right)  $ because the largest of $K$ statistics $D_{\Gamma k}$ has been selected, and it reflects the correlations $\boldsymbol{\rho}_{\Gamma}$ among the coordinates of $\mathbf{D}_{\Gamma}$.

\subsection{Application in the NHANES Example}
\label{ssSubmaxInExample}

Table~\ref{tabResult} performs the test in \S \ref{ssDefineDmax} for the NHANES data in \S \ref{ssIntroEG} using the statistic $T$ of O'Brien and Fleming (1987). The row of Table~\ref{tabResult} for $\Gamma=1$ consists of Normal approximations to randomization tests, while the rows with $\Gamma>1$ examine sensitivity to bias from nonrandom treatment assignment. \ For $\Gamma=1$, the test statistic $D_{\Gamma\max}=6.29\geq\kappa_{\Gamma,\alpha}=2.31$, so Fisher's hypothesis of no treatment effect would be rejected at level $\alpha$ if the data had come from a randomized experiment with $\Gamma=1$. For $\Gamma=1$, the maximum statistic is based on all 470 pairs, $D_{\Gamma\max}=D_{\Gamma1}$; however, $D_{\Gamma k}\geq\kappa_{\Gamma,\alpha}=2.31$ for every subgroup, $k=1,\ldots,K=7$. At $\Gamma=1.4$, the deviates $D_{\Gamma2}$ and $D_{\Gamma6}$ for females ($k=2$) and the nonpoor ($k=6$) no longer exceed $\kappa_{\Gamma,\alpha}=2.31$, and the precise meaning of this is examined in more detail in \S \ref{secClosedTest}. At $\Gamma=1.77$, Fisher's hypothesis of no treatment effect is still rejected because the deviate $D_{\Gamma3}$ for males exceeds $\kappa_{\Gamma,\alpha}=2.31$. Although there are 275 pairs of women and 195 pairs of men, the strongest evidence, the least sensitive evidence, of an effect of inactivity on survival is for men. The bottom of Figure 1 shows the separate survival curves for men and women.

\begin{table}
\small
\caption{\label{tabResult} Seven standardized deviates from Wilcoxon's test, $D_{\Gamma k}$, $k=1,\dots,K=7$, testing the null hypothesis of no effect and their maximum, $D_{\Gamma\mathrm{max}}$,  where the critical value is $d_{\alpha}=  2.31$ for $\alpha=0.05$.  Deviates larger than $d_{\alpha} =  2.31$ are in \textbf{bold}.}
\centering
\begin{tabular}{c|c|cc|cc|cc|c}
\hline
$k$ & 1 & 2 & 3 & 4 & 5 & 6 & 7 \\ \hline
Subpopulation & All & Female & Male & Non-smoker & Smoker & $> 2 \times
$ PL & $\le2 \times$ PL & Maximum \\ \hline& $D_{\Gamma1}$ &  $D_{\Gamma
2}$ & $D_{\Gamma3}$ & $D_{\Gamma4}$ & $D_{\Gamma5}$
& $D_{\Gamma6}$  & $D_{\Gamma7}$ & $D_{\Gamma\mathrm{max}}$ ($p$-values) \\ \hline
Sample-size & 470 & 275 & 195 & 301 & 169 & 254 & 216 \\ \hline
$\Gamma=1.00$ & \textbf{6.29} & \textbf{3.82} & \textbf{5.19} & \textbf{4.84} & \textbf{4.03} & \textbf{3.92} & \textbf{4.96} & \textbf{6.29} (0.000)\\
$\Gamma=1.20$ & \textbf{4.87} & \textbf{2.76} & \textbf{4.24} & \textbf{3.69} & \textbf{3.19} & \textbf{2.93} & \textbf{3.94} & \textbf{4.87} (0.000)\\
$\Gamma=1.40$ & \textbf{3.70} & 1.89 & \textbf{3.47} & \textbf{2.75} & \textbf{2.50} & 2.12 & \textbf{3.11} & \textbf{3.70} (0.001)\\
$\Gamma=1.60$ & \textbf{2.71} & 1.14 & \textbf{2.81} & 1.95 & 1.92 & 1.42 & \textbf{2.40} & \textbf{2.81} (0.013)\\
$\Gamma=1.70$ & 2.26 & 0.80 & \textbf{2.52} & 1.58 & 1.65 & 1.11 & 2.08 & \textbf{2.52} (0.030)\\
$\Gamma=1.77$ & 1.97 & 0.58 & \textbf{2.33} & 1.34 & 1.48 & 0.90 & 1.87 & \textbf{2.33} (0.048)\\
$\Gamma=1.78$ & 1.93 & 0.55 & 2.30 & 1.31 & 1.46 & 0.87 & 1.84 & 2.30 (0.051)\\
\hline
\end{tabular}
\end{table}

Table~\ref{tabResult} is compactly indexed by one parameter $\Gamma$. It can be helpful to give a two-parameter interpretation of $\Gamma$. The longer life of active men in Table~\ref{tabResult} is insensitive to a bias of $\Gamma=1.77$. In a matched pair, $\Gamma=1.77$ corresponds with an unobserved covariate that triples the odds of a longer life and increases the chance of inactivity by a factor of more than 3.5-fold; see the amplification of $\Gamma$ into two parameters $\Delta$ and $\Lambda$ in Rosenbaum (2017, Table 9.1), where $1.77=\Gamma =\left(  \Delta\Lambda+1\right)  /\left(  \Delta + \Lambda\right)$ for $\Delta=3$ and $\Lambda=3.504$.

\subsection{Design Sensitivity and Bahadur Efficiency}

As in Rosenbaum (2012), it is easy to see that under an alternative hypothesis given by a favorable situation --- a treatment effect with no unmeasured bias --- the design sensitivity of $D_{\Gamma\max}$, say $\widetilde{\Gamma}_{\max}$, is equal to the maximum design sensitivity $\widetilde{\Gamma}_{k}$ of the $K$ component tests, $\widetilde{\Gamma}_{\max}=\max\left(  \widetilde{\Gamma}_{1},\ldots,\widetilde{\Gamma}_{K}\right)  $. Briefly, by the definition of design sensitivity, if $\Gamma<$ $\widetilde{\Gamma}_{k}$, then the probability that $D_{\Gamma k}\geq\kappa$ tends to 1 for every $\kappa$ as $\min\left(  I_{g}\right)  \rightarrow\infty$, so the probability that $D_{\Gamma\max}\geq\kappa_{\Gamma,\alpha}$ tends to 1 because $D_{\Gamma\max}\geq D_{\Gamma k}$. Although there is a price to be paid for multiple testing, that price does not affect the design sensitivity.

Define $\beta_{1}=1$. Berk and Jones (1979) show that, if $D_{\Gamma k}$ has Bahadur efficiency $\beta_{k}$ relative to $D_{\Gamma1}$ for $k=2,\ldots,K$, then $D_{\Gamma\max}$ has Bahadur efficiency $\beta_{\max}=\max_{1\leq k\leq K}\beta_{k}$. Berk and Jones call this \textquotedblleft relative optimality\textquotedblright\ meaning $D_{\Gamma\max}$ is optimal among the fixed set $D_{\Gamma1},\ldots,D_{\Gamma K}$. The correction for multiplicity, $\kappa_{\Gamma,\alpha}>\Phi^{-1}\left(  1-\alpha\right)  $, does reduce finite sample power, but in a limited way, so that the Bahadur efficiency is ultimately unaffected.

\subsection{Power Calculations and Design Sensitivity in a Simple Case}
\label{ssPowerCalc}

Under an alternative hypothesis, if the $T_{g}$ are independent and asymptotically Normal with expectation $\mu_{g}^{\ast}$ and variance $\nu_{g}^{\ast}$, then direct manipulations involving the multivariate Normal distribution yield an asymptotic approximation to the power of tests based on $D_{\Gamma\max}$ or $D_{\Gamma k}$.  Write $\theta_{k}^{\ast}=\sum_{g=1}^{G}c_{gk}\,\mu_{g}^{\ast}$ and $\sigma_{k}^{\ast}$ for the square root of the $k$th diagonal element of $\mathbf{C}\mathrm{diag}\left(  \nu_{1}^{\ast},\ldots,\nu_{K}^{\ast}\right)  \mathbf{C}^{T}$, so $\theta_{k}^{\ast}$ is the expectation and $\sigma_{k}^{\ast}$ is the standard deviation of $S_{k}$ under the alternative. Write $\boldsymbol{\rho}^{\ast}$ for the correlation matrix computed from this covariance matrix. The approximate power is $1-\Pr\left(  D_{\Gamma\max}<\kappa_{\Gamma,\alpha}\right)$, where $\Pr\left(  D_{\Gamma\max}<\kappa_{\Gamma,\alpha}\right)$ is:
\[
\Pr\left( \frac{S_{k}-\theta_{\Gamma k}}{\sigma_{\Gamma k}}<\kappa_{\Gamma,\alpha},\,k=1,\ldots,K\right)=\Pr\left(  \frac{S_{k}-\theta_{k}^{\ast}}{\sigma_{k}^{\ast}}<\frac{\theta_{\Gamma k}-\theta_{k}^{\ast}+\kappa_{\Gamma,\alpha}\,\sigma_{\Gamma k}}{\sigma_{k}^{\ast}},\,k=1,\ldots,K\right)  \text{.}
\]
So, $\Pr\left(  D_{\Gamma\max}<\kappa_{\Gamma,\alpha}\right)$ is approximately a particular quadrant probability for the $N_{K}\left(  \mathbf{0},\boldsymbol{\rho}^{\ast}\right)$ distribution, and this may be calculated using the \texttt{pmvnorm} function in the \texttt{mvtnorm} package in \texttt{R}. Under the same assumptions, the power of a test based on one fixed $D_{\Gamma k}$ is approximately
\begin{equation}
1-\Pr\left\{  \frac{S_{k}-\theta_{k}^{\ast}}{\sigma_{k}^{\ast}}<\frac{\theta_{\Gamma k}-\theta_{k}^{\ast}+\Phi^{-1}\left(  1-\alpha\right) \,\sigma_{\Gamma k}}{\sigma_{k}^{\ast}}\right\}  \text{,} \label{eqPowerOneDk}
\end{equation}
and this may be calculated using the standard Normal distribution.

Moreover, the design sensitivity $\widetilde{\Gamma}_{k}$ for $S_{k}=\sum_{g=1}^{G}c_{gk}T_{g}$ is the limit of values of $\Gamma$ that solve $1=\left(  \sum_{g=1}^{G}c_{gk}\,\mu_{g}^{\ast}\right)  /\left(  \sum_{g=1}^{G}c_{gk}\,\mu_{\Gamma g}\right)  $. That is, using $S_{k}$, as $I\rightarrow\infty$, the power tends to 1 for $\Gamma<\widetilde{\Gamma}_{k}$ and it tends to 0 for $\Gamma>\widetilde{\Gamma}_{k}$. This formula emphasizes the importance of effect modification. For instance, with two groups, $G=2$, say $g=0$ and $g=1$, if $\mu_{0}^{\ast}>\mu_{1}^{\ast}$, then the design sensitivity is largest with $c_{0k}=1$ and $c_{1k}=0$, so as $I\rightarrow\infty$, there are values of $\Gamma$ such that the power of the overall test is tending to 0 while the power of a test focused on the first subgroup is tending to 1. This will be quite visible in power calculations.

An oracle uses the one $D_{\Gamma k}$ with the highest power. Lacking an oracle, it is interesting to compare $D_{\Gamma\max}$ to: (i) the oracle, (ii) the test, $D_{\Gamma1}$, that uses all of the matched sets.

To illustrate, consider the simple, balanced case with $I_{g}=I_{+}/G=\overline{I}$, say, for every $g$, and suppose that there are $L$ binary covariates as potential effect modifiers. We would like to compute power under a favorable alternative, meaning that, unknown to the investigator, the treatment has an effect and there is no unmeasured bias from $u_{gij}$. Because the investigator cannot know that the data came from the favorable situation, a sensitivity analysis is performed. A simple favorable situation has $\overline{I}$ independent treated-minus-control pair differences in every group $g$, where the pair differences are Normal with various expectations and variance 1. Then Wilcoxon's signed rank statistic in group $g$, namely $T_{g}$, is asymptotically Normal under the alternative hypothesis as $\overline{I}\rightarrow\infty$, and simple formulas in Lehmann (1975, \S 4.2) give the expectation and variance, $\mu_{g}^{\ast}$ and $\nu_{g}^{\ast}$, of $T_{g}$, under this alternative. In this case, $\mu_{\Gamma g}$ and $\nu_{\Gamma g}$ are given in Rosenbaum (2002, \S 4.3.3).  There are $G\overline{I}=2^{L}\cdot\overline{I}$ pairs in total. Note that the $K=2L+1$ statistics, $S_{k}$, are each computed from at least $2^{L-1}\cdot\overline{I}$ pairs, not from $\overline{I}$ pairs, and they are each sums of at least $2^{L-1}$ signed rank statistics $T_{g}$.  If $L=3$ in this balanced design, then, under $H_{0}$, two different levels, say men and women, of one potential effect modifier, gender, have uncorrelated $S_{k}$, two levels of different effect modifiers have $S_{k}$ with correlation 0.5, and the overall statistic, $S_{1}$, has correlation 0.707 with each component test, $S_{k}$ for $k=2,\dots,7$, so most pairs of test statistics are strongly correlated.  Wilcoxon's test is familiar and convenient for a power calculation in this balanced design with $I_{g}=\overline{I}$; however, unlike an M-test or the test of O'Brien and Fleming (1987), Wilcoxon's signed rank test would need rescaling before summing over $g$ in an unbalanced design.

\begin{table}
\caption{\label{tabPower} Theoretical power for Wilcoxon's signed rank test in subgroup analyses using (i) the maximum statistic $D_{\Gamma\max}$, (ii) an oracle that knows a priori which group has the largest effect (Oracle), and (iii) one statistic that sums all Wilcoxon statistics, thereby using all the matched pairs, $D_{\Gamma1}$.  }
\centering
\begin{tabular}{l|l|rrr|rrr}
\hline Situation & & \multicolumn{3}{|c}{One covariate, $L=1$} & \multicolumn
{3}{|c}{Five covariates, $L=5$} \\
\hline
& $\Gamma$ & $D_{\Gamma\max}$ & Oracle & $D_{\Gamma1}$ & $D_{\Gamma\max
}$ & Oracle & $D_{\Gamma1}$\\
\hline$ (\zeta_0, \zeta
_1)=(0,0)$ & 1 & 0.050 & 0.050 & 0.050 & 0.050 & 0.050 & 0.050\\
1. No effect.  Values  & 1.01 & 0.035 & 0.033 & 0.033 & 0.035 & 0.033 & 0.033\\
are the size test. & 1.1 & 0.000 & 0.000 & 0.000 & 0.000 & 0.000 & 0.000\\
& 1.3 & 0.000 & 0.000 & 0.000 & 0.000 & 0.000 & 0.000\\ \hline$ (\zeta
_0, \zeta_1)=(0.5, 0.5)$ & 1 & 1.000 & 1.000 & 1.000 & 1.000 & 1.000 & 1.000\\
2. Constant effect. & 2.8 & 0.579 & 0.671 & 0.671 & 0.460 & 0.601 & 0.601\\
Every subgroup & 3.0 & 0.177 & 0.218 & 0.218 & 0.126 & 0.167 & 0.167\\
has effect 0.5. & 3.2 & 0.030 & 0.030 & 0.030 & 0.020 & 0.019 & 0.019\\
& 3.4 & 0.004 & 0.002 & 0.002 & 0.002 & 0.001 & 0.001\\ \hline$ (\zeta
_0, \zeta_1)=(0.6, 0.4)$ & 1 & 1.000 & 1.000 & 1.000 & 1.000 & 1.000 & 1.000\\
3. Moderate effect & 2.8 & 0.991 & 0.998 & 0.593 & 0.959 & 0.996 & 0.521\\
modification, & 3.0 & 0.928 & 0.971 & 0.161 & 0.791 & 0.959 & 0.121\\
$\zeta_0 > \zeta_1$& 3.2 & 0.733 & 0.855 & 0.018 & 0.492 & 0.816 & 0.011\\
& 3.4 & 0.446 & 0.615 & 0.001 & 0.220 & 0.554 & 0.001\\
\hline
\end{tabular}
\end{table}

Table~\ref{tabPower} displays theoretical power for a level $\alpha=0.05$ test of no effect in several favorable situations, that is, situations with a treatment effect and no bias. In Table~\ref{tabPower}, \textquotedblleft one covariate\textquotedblright\ refers to $L=1$ binary covariate, making $G=2^{L}=2$ groups, so that $D_{\Gamma\max}$ is the maximum of three statistics, namely the deviates for the signed rank statistics in groups 1 and 2 and for the sum of these two statistics. In Table~\ref{tabPower}, \textquotedblleft five covariates\textquotedblright\ refers to $L=5$ binary covariates, making $G=2^{L}=32$ groups, so that $D_{\Gamma\max}$ is the maximum of $11=2\times5+1$ statistics, namely the deviates for 10 totals of 16 signed rank statistics at the high and low levels of each covariate, and also for the sum of all 32 signed rank statistics.

The sample size in Table~\ref{tabPower} is constant, $I_{g}=\overline{I}$, with total   $2016=G\overline{I}=2^{L}\cdot\overline{I}$, so $\overline{I}=1008$ for $L=1$ covariate and $\overline{I}=63$ for $L=5$ covariates. In both cases, $L=1$ and $L=5$, only the first covariate is a potential effect modifier: the expected pair difference only changes with the level of the first covariate, being $\zeta_{0}$ for the 0 level and $\zeta_{1}$ for the 1 level. When $\zeta_{0}\neq\zeta_{1}$, there is effect modification. With $L=5$, four of five covariates are distractions requiring a larger correction for multiple testing. The first situation in Table~\ref{tabPower} has no effect, $\zeta_{0}=\zeta_{1}=0$, so the values are the actual size of a level $\alpha=0.05$ test. The second situation in Table~\ref{tabPower} has a constant treatment effect, $\zeta_{0}=\zeta_{1}=0.5$, so it is a mistake to look for effect modification because there is none. The third situation in Table~\ref{tabPower} has moderate effect modification, $\zeta_{0}=0.6>0.4=\zeta_{1}$, but the average effect is still $0.5=\left(  \zeta_{0}+\zeta_{1}\right)/2$.

Table~\ref{tabPower} compares the power of $D_{\Gamma\max}$ to a single combined test $D_{\Gamma1}$ that uses all pairs and an oracle that performs a single test using all the pairs that have the largest value of $\zeta_{g}$. Obviously, the oracle is not a statistical procedure because it requires the statistician to know what she does not know, namely which groups have the largest $\zeta_{g}$. From theory, in the nonnull situations 2 and 3, we know that $D_{\Gamma\max}$ has the same design sensitivity as the oracle, whereas  $D_{\Gamma1}$ has lower design sensitivity than the oracle unless there is no effect modification, $\zeta_{0}=\zeta_{1}$, as in situation 2. In situation 2, all three procedures have design sensitivity $\widetilde{\Gamma}=3.17$, with negligible power for $\Gamma=3.2>3.17$. In situation 3, $\zeta_{0}=0.6$, and both $D_{\Gamma\max}$ and the oracle have design sensitivity $\widetilde{\Gamma}=4.05$ by focusing on group 0 for covariate 1, and they have nonnegligible power at $\Gamma=3.4<4.05$; however, $D_{\Gamma1}$ has design sensitivity $\widetilde{\Gamma}=3.13$ in situation 3, with negligible power at $\Gamma=3.2$.

In the first situation in Table~\ref{tabPower}, all tests have the correct size for $\Gamma=1$, and because there is no actual bias in the favorable situation, they have size below 0.05 for $\Gamma>1$. In the second situation in Table~\ref{tabPower}, $D_{\Gamma\max}$ pays a price, searching for effect modification that is not there. In situation 3, $D_{\Gamma\max}$ has much higher power than $D_{\Gamma1}$, but it is behind the oracle, reflecting the price paid to discover the true pattern of effect modification. For instance, at $\Gamma=2.8$, with $L=5$ binary covariates and moderate effect modification, $\zeta_{0}=0.6>0.4=\zeta_{1}$, the statistic $D_{\Gamma\max}$ has power 0.959, the oracle has power 0.996, and $D_{\Gamma1}$ has power 0.521. Simulated power is discussed in an on-line appendix.

\section{Simultaneous Inference and Closed Testing}
\label{secClosedTest}

Strictly speaking, the statistic $D_{\Gamma\max}$ is a test of a global null hypothesis, specifically Fisher's hypothesis $H_{0}$ of no treatment effect in the study as a whole. In previous sections, the $c_{gk}$ are either 0 or 1, and the $k$th comparison defines a subpopulation $\mathcal{S}_{k}$ as those groups with $c_{gk}=1$, that is, $\mathcal{S}_{k}=\left\{  g:c_{gk}=1\right\}$, for instance, the subpopulation of men. We are, of course, interested in the hypothesis, say $H_{k}$, that asserts there is no effect in subpopulation $\mathcal{S}_{k}$, say no effect in the subpopulation of men. We would like to test all $K$ hypotheses $H_{k}$, $k=1,\ldots,K$, strongly controlling the family-wise error rate at $\alpha$ in the presence of a bias of at most $\Gamma$. We may do this with the closed testing method of Marcus et al. (1976).

Define $H_{\mathcal{I}}$ for $\mathcal{I}\subseteq\left\{  1,\ldots,K\right\}$ to be the hypothesis of no treatment effect in the union of the subpopulations $\mathcal{S}_{k}$, $k\in\mathcal{I}$. Then $H_{\left\{  2,5\right\}  }$ says that there is no effect for females, $k=2$, and for smokers, $k=5$. If $H_{\left\{2,5\right\}  }$ were true, there might be an effect for male nonsmokers. If the goal were to test $H_{\mathcal{I}}$ at level $\alpha$ in the presence of a bias of at most $\Gamma$, then this could be done using $D_{\Gamma\mathcal{I}}=\max_{k\in\mathcal{I}}D_{\Gamma k}$, which is a test of the same form as $D_{\Gamma\max}$, whose approximate critical constant from (\ref{eqCritConst}), say $\kappa_{\Gamma,\alpha,\mathcal{I}}$, must be calculated using a $\left\vert \mathcal{I}\right\vert $-dimensional Normal distribution. Of course, $D_{\Gamma\mathcal{I}}\geq D_{\Gamma\mathcal{J}}$ whenever $\mathcal{J}\subset\mathcal{I}$, so $\kappa_{\Gamma,\alpha,\mathcal{J}}\leq\kappa_{\Gamma,\alpha,\mathcal{I}}$; that is, the correction for multiple testing is less severe when fewer comparisons are made. In particular, $\kappa_{\Gamma,\alpha,\mathcal{I}}\leq\kappa_{\Gamma,\alpha}$ for all $\mathcal{I}\subseteq\left\{1,\ldots,K\right\}$.

The closed testing method of Marcus et al. (1976) rejects $H_{\mathcal{I}}$ at level $\alpha$ in the presence of a bias of at most $\Gamma$ if it rejects $H_{\mathcal{K}}$ for all $\mathcal{K}\supseteq\mathcal{I}$, that is, if $D_{\Gamma\mathcal{K}}\geq\kappa_{\Gamma,\alpha,\mathcal{K}}$ for all hypotheses $\mathcal{K}$ that contain $\mathcal{I}$. Closed testing has attractive properties. Closed testing strongly controls the family-wise error rate, as demonstrated by Marcus et al. (1976). This property extends to sensitivity analyses. No matter which hypotheses are true or false, the probability that closed testing rejects at least one true $H_{\mathcal{I}}$ is at most $\alpha$ if the bias is at most $\Gamma$.  In contrast, use of the Bonferroni inequality in sensitivity analysis is somewhat conservative; see Rosenbaum and Silber (2009, \S 4.4-\S 4.5) and Fogarty and Small (2016).

To illustrate, consider $\Gamma=1.4$ in Table~\ref{tabResult}, where the deviates for females ($k=2$) and for nonpoor ($k=6$) would not have led to rejection of the global null hypothesis $H_{0}$ of no effect.  At $\Gamma=1.4$, closed testing rejects the hypothesis of no effect in each of the six subgroups in Table~\ref{tabResult}, including females and the nonpoor.  When closed testing tests $H_{2,6}$, the hypothesis asserting no effect for women and for the nonpoor, the critical value is no longer $\kappa_{\Gamma,\alpha }=2.31$ but rather $\kappa_{\Gamma,\alpha,\left\{2,6\right\}  }=1.92$, leading to rejection at $\alpha = 0.05$ in the presence of a bias of at most $\Gamma=1.4$.  Because of this rejection, closed testing continues on to test $H_{2}$ with revised critical value $\kappa_{\Gamma,\alpha,\left\{2 \right\}  }=1.65$, leading to rejection of no effect for females.

When converting a global test into a closed testing procedure, one must ensure that the assumptions of the global test are satisfied when testing each component hypothesis, $H_{\mathcal{I}}$.  In particular, the scores, $q_{gij}$ must be functions of $\mathcal{F}$ when $H_{\mathcal{I}}$ is true; see \S\ref{ssNotationCausal}.  This happens if $q_{gij}$ is a function of responses $R_{gij}$ in group $g$ for each $g$, as in the example, where the Prentice-Wilcoxon scores were computed separately in each of the $2^{L} = 2^{3} = 8$ interaction groups $g$.  More generally, a simple rule says: the $q_{gij}$ used to test $H_{\mathcal{I}}$ can depend upon $R_{gij}$ only if $c_{gk} = 1$ for at least one $k \in \mathcal{I}$.  See the documentation for the \texttt{score} function in the \texttt{submax} package in \texttt{R} for further discussion.

It is possible to strengthen closed testing when there are logical implications among the hypotheses, $H_{1},\ldots,H_{K}$, as is true here. Strengthening changes the procedure so that it still controls the family-wise error rate but it may, from time to time, reject an additional hypothesis not rejected by closed testing. Holm's method is the application of closed testing using the Bonferroni inequality, and Shaffer (1986) strengthened Holm's method when applied to the analysis of variance using logical implications among hypotheses. What are the logical implications in Table~\ref{tabResult}? Recall that hypotheses assert that no one in certain subpopulations was affected by the treatment. If any of $H_{2},\ldots,H_{K}$ is false, then $H_{1}$ is false. Similarly, if $H_{5}$ is false, so at least some smokers are affected, then either $H_{2}$ or $H_{3}$ or both must be false, because every smoker is either male or female. Bergmann and Hommel (1988) discuss the steps required to strengthen a closed testing procedure based on logical implications among hypotheses.  A related strategy is discussed by Goeman and Finos (2012).  In principle, a closed testing or stepwise testing procedure may be inverted to obtain confidence sets; see Hayter and Hsu (1994) for discussion.

\section{Pairs or Sets That Are Not Exactly Matched for Some Effect Modifiers}
\label{secInexact}

To avoid confusing a main effect of gender and effect modification involving gender, we search for effect modification by gender in sets that are exactly matched for gender, say in pairs of women. In the example in \S \ref{ssIntroEG}, all pairs were exactly matched for gender, smoking and the indicator of an income above twice the poverty level. Sometimes, it may not be possible to match exactly for every potential effect modifier. What can be done in this case?  The procedure is direct, but it requires some additional bookkeeping.  We keep track of inexactly matched pairs and make a change in the comparison weights $c_{gk}$.  However, we do not increase the number of tests, $K$.  We use a pair of women in the comparison for women even if that pair is not exactly matched for income or smoking.

Suppose that exact matching for $L$ binary effect modifiers is not possible.  So-called ``almost-exact matching'' tolerates some inexact matches but minimizes their number; see Rosenbaum (2010, \S9.2).  Typically, the matching would balance all covariates even when they are not exactly matched, perhaps by also matching on the propensity score, so inexact matching would not, by itself, introduce confounding.  Instead of $G=2^{L}$ groups of exactly matched pairs, there would be $G=2^{L} \times 2^{L} = 2^{2L}$ groups of pairs for the different ways the $L$ effect modifiers might be matched or mismatched.  For example, one group $g$ consists of pairs of nonsmoking women in which the treated woman is poor and the control is not poor.  That group of pairs would be included in the comparison for women, and also in the comparison for nonsmokers, but would not be included in the two comparisons for poor and for not poor.

Now $G=2^{2L}$, instead of $G=2^{L}$, so the definition of $c_{gk}$ changes.  If comparison $k$ refers to women, then $c_{gk}=1$ if group $g$ contains pairs of women, and otherwise $c_{gk}=0$.  That is, $c_{gk}=0$ if the pairs in group $g$ contain either one or two men.  The statistic $D_{\Gamma k}$ then refers to all pairs of two women, whether or not smoking and poverty are exactly matched.  Importantly, the number of groups, $G=2^{2L}$, has increased but the number of tests, $K$, has not increased.

The on-line appendix simulates the proportion of pairs
that are exactly matched for one effect modifier.  Because we care about exact matches for one effect modifier at a time, not for all at once, this proportion is quite high.

\section{Discussion}

Effect modification is important in observational studies for several reasons.

With effect modification, we expect to report firmer causal conclusions in subpopulations with larger effects. That is, we expect the design sensitivity and the power of the sensitivity analysis to be larger, so we expect to report findings that are insensitive to larger unmeasured biases in these subpopulations. Such a discovery is important in three ways. First, the finding about the affected subpopulation is typically important in its own right as a description of that subpopulation. Second, if there is no evidence of an effect in the complementary subpopulation, then that may be news as well. Third, if a sensitivity analysis convinces us that the treatment does indeed cause effects in one subpopulation, then this fact demonstrates the treatment does sometimes cause effects, and it makes it somewhat more plausible that smaller and more sensitive effects in other subpopulations are causal and not spurious. This is analogous to the situation in which we discover that heavy smoking causes lots of lung cancer, and are then more easily convinced that second-hand smoke causes some lung cancer, even though the latter effect is much smaller and more sensitive to unmeasured bias.

Conversely, it can be useful to discover evidence of a treatment effect of the same sign in every major subpopulation. We often worry whether findings generalize to another population that was not studied. Will a study done in Georgia generalize to Kansas where no study was done? If the second population were simply a different mixture of the same types of people --- e.g., in Table~\ref{tabResult}, a different mixture of men and women, smokers and nonsmokers, rich and poor --- then finding strong evidence of a nontrivial effect of constant sign in all subpopulations provides reason to think that the direction of effect will reappear in the second population.

How many potential effect modifiers should be examined?  With $L$ potential binary effect modifiers, $2L+1$ correlated tests are performed.  The proposed method corrects, as it must, for testing several hypotheses.  There is a trade-off between the severity of this correction for multiple testing and the possibility of failing to examine, hence failing to locate, an important effect modifier.  The loss of power due to testing $L=5$ potential effect modifiers when only one of these is actually an effect modifier is quantified in Table~\ref{tabPower} and the on-line appendix, and similar calculations may be performed for other values of $L$ and $I$, and for other distributions.  It is difficult to offer advice applicable in all scientific contexts, except for the following observations. First, one can err in both directions, either setting $L$ high and paying a high price for multiple testing, or setting $L$ low and missing an important effect modifier.  Second, the power of the sensitivity analysis is affected by both $L$ and the sample sizes, $I_{g}$, so power calculations using the actual $I_{g}$ may be helpful.

The simulation in the on-line appendix contrasted the new submax method with another method using groups formed by CART. One difference between the two methods is that there is more theory concerning the performance of the submax method, including power, design sensitivity and Bahadur efficiency. The submax method achieves the largest design sensitivity of the subgroups, but there is no similar claim for the CART method. In the simulation,  CART was cautious about forming groups, so it failed to capitalize on moderate effect modification, with a loss of power in some situations; however, that also meant that CART rarely paid a price for multiple testing when there was no effect modification.

The submax and CART methods may be combined in several ways.  For instance, an investigator may combine a few potential effect modifiers selected a priori with a few groups suggested by CART, applying the submax method to all of these groups.

%

\pagebreak
\thispagestyle{plain}

\begin{center}
\textbf{\Large A Powerful Approach to the Study of Moderate Effect Modification in Observational Studies: On-Line Appendix}
\end{center}

\vspace{0.3cm}

\begin{center}
\textbf{Kwonsang Lee$^{1}$, Dylan S. Small$^{2}$, and Paul R. Rosenbaum$^{2}$} \\[0.3cm]
{\small $^{1}$Department of Biostatistics, Harvard School of Public Health\\
$^{2}$Department of Statistics, University of Pennsylvania}
\end{center}

\setcounter{equation}{0}
\setcounter{figure}{0}
\setcounter{table}{0}
\setcounter{page}{1}
\makeatletter
\renewcommand{\theequation}{S\arabic{equation}}
\renewcommand{\thefigure}{S\arabic{figure}}

\begin{abstract}
This on-line appendix addresses several issues.  The simulated power of the submax method is compared with a method based on CART.  The CART method is illustrated in the example.  The loss of sample size, and hence the loss of power, due to inexact matching for effect modifiers is examined in simulation.

\end{abstract}

\subsection*{3.6 Simulated Power and a Comparison with CART Groups}
\label{ssSimulation}

Appendix Table~\ref{tabSim} describes simulated power for some of the same situations as the theoretical power in Table 3 of the paper. The simulation includes a competing method for matched pairs proposed by Hsu et al. (2015), in which groups are built from covariates using a CART procedure. There is no known power formula for the CART method, so it cannot be included in Table 3 of the paper. In this approach, the pairs are initially ungrouped, and so lack a $g$ subscript. However, the pairs have been exactly matched for several covariates that may be effect modifiers. The absolute treated-minus-control pair difference in outcomes in pair $i$, namely $\left\vert Y_{i}\right\vert=\left\vert R_{i1}-R_{i2}\right\vert $, is regressed on these covariates using CART, and the leaves of the tree define the groups. The $P$-values with the groups so-defined are combined using the truncated product of $P$-values proposed by Zaykin et al. (2002). The truncated product is analogous to Fisher's product of $P$-values, except $P$-values above a prespecified truncation point, $\varsigma$, enter the product as 1, so the two methods are the same for $\varsigma=1$. In  Table 3 of the paper, $\varsigma=1/10$. Unlike $D_{\Gamma\max}$, there is no guarantee that the CART procedure will equal the oracle in terms of design sensitivity. In other words, we expect $D_{\Gamma\max}$ to win in sufficiently large samples, tracking the oracle as $\min\left(  I_{g}\right)  \rightarrow\infty$; however, $D_{\Gamma\max}$ may not win in the finite samples.

The CART method makes discrete choices: whether to create subgroups, which groups to create. We expect the CART method to perform well when it makes correct choices, so we expect it to perform well in extreme situations in which the correct choices are fairly clear: no effect modification, or dramatic effect modification. In Appendix Table \ref{tabSim}, the CART method is close to the oracle when there is no effect modification, and it is substantially inferior to both the submax method and the oracle when there is moderate effect modification.

\begin{table}
\caption{\label{tabSim} Simulated power (number of rejections in 10,000 replications) for Wilcoxon's signed rank test in subgroup analyses using (i) the maximum statistic $D_{\Gamma\max}$, (ii) groups built by CART, (iii) an oracle that knows a priori which group has the largest effect (Oracle), and (iv) one statistic that sums all of the Wilcoxon statistics, thereby using all matched pairs, $D_{\Gamma1}$.}
\centering
\begin{tabular}{l|ll|rrrr|rrrr}
\hline
& & & \multicolumn{4}{|c}{One covariate, $L=1$} & \multicolumn{4}%
{|c}{Five covariates, $L=5$} \\
\hline
& $\zeta= (\zeta_0, \zeta_1)$ & $\Gamma$ & $D_{\Gamma\max}%
$ & CART & Oracle & $D_{\Gamma1}$ & $D_{\Gamma\max}%
$ & CART & Oracle & $D_{\Gamma1}$\\
\hline
1 & (0,0) & 1 & 540 & 525 & 525 & 525 & 515 & 504 & 503 & 503 \\
& & 1.1 & 7 & 1 & 1 & 1 & 7 & 7 & 7 & 7\\[0.3cm]
2 & (0.5, 0.5) & 1 & 10000 & 10000 & 10000 & 10000 & 10000 & 10000 & 10000 & 10000 \\
& & 2.8 & 5804 & 6713 & 6713 & 6713 & 4581 & 6014 & 6014 & 6014\\
& & 3.0 & 1643 & 2104 & 2101 & 2101 & 1215 & 1685 & 1681 & 1681\\[0.3cm]

3 & (0.55, 0.45) & 1 & 10000 & 10000 & 10000 & 10000 & 10000 & 10000 & 10000 & 10000 \\
& & 2.8 & 8263 & 6618 & 9035 & 6541 & 6729 & 5905 & 8769 & 5814 \\
& & 3.0 & 5011 & 2178 & 6549 & 2030 & 2900 & 1673 & 6035 & 1520\\
& & 3.2 & 1980 & 307 & 3412 & 215 & 795 & 272 & 2927 & 166\\
& & 3.4 & 521 & 47 & 1190 & 20 & 166 & 46 & 976 & 7\\[0.3cm]

4 & (0.6, 0.4) & 1 & 10000 & 10000 & 10000 & 10000 & 10000 & 10000 & 10000 & 10000\\
& & 2.8 & 9913 & 7073 & 9977 & 6058 & 9589 & 6584 & 9955 & 5348\\
& & 3.0 & 9264 & 3788 & 9701 & 1657 & 7975 & 3471 & 9588 & 1242 \\
& & 3.2 & 7387 & 2313 & 8565 & 173 & 5071 & 2212 & 8208 & 121\\
& & 3.4 & 4603 & 1535 & 6265 & 6 & 2245 & 1363 & 5679 & 8\\[0.3cm]

5 & (0.65, 0.35) & 1 & 10000 & 10000 & 10000 & 10000 & 10000 & 10000 & 10000 & 10000 \\
& & 3.0 & 9978 & 7602 & 9992 & 968 & 9862 & 7548 & 9996 & 729\\
& & 3.3 & 9524 & 7090 & 9811 & 17 & 8492 & 7045 & 9758 & 6\\
& & 3.6 & 7283 & 5682 & 8470 & 0 & 4857 & 5391 & 8086 & 0\\
& & 3.9 & 3564 & 2967 & 5329 & 0 & 1594 & 2586 & 4659 & 0\\[0.3cm]
\hline
\end{tabular}
\end{table}

In the first situation in Appendix Table 1, there is no treatment effect. All four methods falsely reject the null hypothesis of no treatment effect about five percent of the time when $\Gamma=1$, there is no effect, and the nominal level is 0.05. Appendix Table 1 checks the theoretical formulas that yielded Table 3 in the paper, and in general the two tables are in agreement. The CART procedure has higher power than\ $D_{\Gamma\max}$ when there is no effect modification in situation 2, $\zeta_{0}=\zeta_{1}=0.5$, because it typically produces a single group in this situation. The CART procedure has lower power than\ $D_{\Gamma\max}$ when there is moderate effect modification in situation 3, $\zeta_{0}=0.55>0.45=\zeta_{1}$, perhaps because the CART procedure fails to locate the moderate effect modification. In situation 5, with $\zeta_{0}=0.65>0.35=\zeta_{1}$, the submax method has higher power than the CART method with $L=1$ covariate and with $L=5$ covariates for $\Gamma \le 3.3$, but the CART method has higher power with $L=5$ covariates and $\Gamma \ge 3.6$. In Appendix Table \ref{tabSim}, using all of the data in a single test is inferior except when there is no effect modification at all. The submax method performs well when there is moderate effect modification.

\subsection*{3.7 Use of CART in the Example}
\label{ssCARTinExample}

As an alternative method, consider using the CART method in \S \ref{ssSimulation}. Using the default settings in \texttt{rpart} in \texttt{R}, the CART tree is a single group of all 470 pairs. At $\Gamma=1.77$, the single group test has deviate $D_{\Gamma1}=1.97$ and one-sided $P$-value bound of $1-\Phi\left(1.97\right)  =0.024$. If the complexity parameter in \texttt{rpart} is reduced below 0.0062, then the CART tree splits on sex. Hsu et al. combine  $P$-value bounds from leaves of the tree using Zaykin et al. (2002)'s truncated product, as in \S\ref{ssSimulation}. At $\Gamma=1.77$, if the two $P$-value bounds for females and males, $1-\Phi\left(  0.58\right)  =0.281$ and $1-\Phi\left(2.33\right)  =0.010$, are combined, then the combined $P$-value bound is 0.028. In this one example, the two analyses give similar impressions.

\subsection*{5.1 Loss of pairs due to inexact matching for effect modifiers}

As discussed in \S5 of the paper, a pair that is inexactly matched for an
effect modifier is not used in the test for that effect modifier, but this
same pair is used for other effect modifiers for which the pair is exactly
matched.  A pair consisting of a male smoker and a female smoker is not
used when studying effect modification by gender, but is used when studying
effect modification by smoking.  When some pairs are not used in this way,
there will be a consequent loss of power in the study of effect modification
by gender.  How many pairs are lost in this way?

It is important to keep in mind that we do not require exact matching for
all effect modifiers at once, but only exact matching for the one effect
modifier currently being studied.  If we were matching for, say, 5
effect modifiers, then half the pairs might be inexactly matched for at
least one of the 5 effect modifiers, yet each effect modifier might be
exactly matched for 90\% of the pairs, because $50\% = 5 \times 10\%$.
The loss of power in this case would be small.

To illustrate, we consider a simple situation in which $p$ binary
effect modifiers are constructed by cutting a $p$ dimensional Normal distribution,
so it is a form of probit model.  The Normal variables were cut at zero to
form $p$ binary variables.  We consider $p=3$, 5, 7.  In the control
population, the $p$ Normal random variables have expectation zero.
In the treated population, the $p$ Normal random variables each
have expectation $\zeta/p$, so the total of the $p$ variables has expectation
$\zeta$ for every $p$.  The $p$ Normal random variables all have correlation
$\xi$ with one another in both treated and control populations.  Because
the $p$ variables are exchangeable, we may report results for any one of them.
(Actually, we determined the simulated results for the coordinates one at a
time, and averaged the $p$ results, thereby decreasing the standard error
of the expected result for one coordinate.)  In all cases, $n_{t}$ pairs
were formed from $n_{t}$ treated subjects and $2 \times n_{t}$ potential
controls.  This is a smaller, less favorable matching ratio than in the NHANES
example where there were $3.15 \times n_{t}$ potential controls available
to form matched pairs.  We consider both $n_{t}=500$ pairs, 
analogous to the NHANES example, and $n_{t}=1000$ pairs.  Pairs
were optimally matched for the $p$ potential effect modifiers.  Each sampling
situation was replicated 1000 times, but the reported proportion of exact
matches for one effect modifier is an average of $1000 \times p$ sample
proportions.

\begin{table}[h]
\centering
\caption{Mean proportion of exact matches for one effect modifier.}
\label{tabInexact}
\begin{tabular}{ccccc}
\hline
Correlation & Expected & \multicolumn{3}{c}{Number of effect } \\
 & Total & \multicolumn{3}{c}{ modifiers $p$} \\
$\xi$ & $\zeta$ & 3 & 5 & 7 \\
\hline
\multicolumn{5}{c}{$n_{t}=500$ pairs.} \\
\hline
0 & 0 & 1.000 & 1.000 & 0.988 \\
& 0.5 & 1.000 & 0.999 & 0.986 \\
& 1 & 0.996 & 0.993 & 0.975 \\
& 2 & 0.892 & 0.860 & 0.891 \\
0.25 & 0 & 1.000 & 0.999 & 0.987 \\
& 0.5 & 1.000 & 0.999 & 0.987 \\
& 1 & 1.000 & 0.999 & 0.987 \\
& 2 & 0.948 & 0.969 & 0.973 \\
\hline
\multicolumn{5}{c}{$n_{t}=1000$ pairs.} \\
\hline
0 & 0 & 1.000 & 1.000 & 0.997 \\
& 0.5 & 1.000 & 1.000 & 0.996 \\
& 1 & 0.997 & 0.997 & 0.990 \\
& 2 & 0.898 & 0.872 & 0.905 \\
0.25 & 0 & 1.000 & 1.000 & 0.997 \\
& 0.5 & 1.000 & 1.000 & 0.997 \\
& 1 & 1.000 & 1.000 & 0.996 \\
& 2 & 0.953 & 0.979 & 0.986 \\
\hline
\end{tabular}
\end{table}

Appendix Table \ref{tabInexact} shows the simulated proportions of
exact matches for one effect modifier in the presence of
$p$ effect modifiers.  Except when the bias $\zeta$ is 
very large, $\zeta=2$, the proportion of exact matches
is close to 1, and even for $\zeta=2$ it is always at 
least 85\%.  For the situations in Appendix Table \ref{tabInexact},
the loss of power due to inexact matching is not likely to
be large.


\begin{thebibliography}{plain}

\bibitem{bergmann1988}
Bergmann, B., Hommel, G. (1988). Improvements of general multiple test procedures for redundant systems of hypotheses. {\it Multiple Hypothesenpr\"{u}fung}, NY: Springer, 100--115.

\bibitem{berk1978}
Berk, R. H. and Jones, D. H. (1978). Relatively optimal combinations of test statistics. {\it Scandinavian Journal of Statistics} {\bf 5}, 158--162.

\bibitem{davis1994}
Davis, M. A., Neuhaus, J. M., Moritz, D. J., Lein, D., Barclay, J. D. and Murphy, S. P. (1994). Health behaviors and survival among middle aged and older men and women in the NHANES I Epidemiologic Follow-Up Study. {\it Preventive Medicine} {\bf 23}, 369--376.

\bibitem{egleston2009}
Egleston, B. L., Scharfstein, D. O. and MacKenzie, E. (2009). On estimation of the survivor average causal effect in observational studies when important confounders are missing due to death. {\it Biometrics} {\bf 65}, 497--504.

\bibitem{fogarty2016}
Fogarty, C. B. and Small, D. S. (2016). Sensitivity analysis for multiple comparisons in matched observational studies through quadratically constrained linear programming. {\it Journal of the American Statistical Association} {\bf 111}, 1820--1830.

\bibitem{gastwirth2000}
Gastwirth, J. L., Krieger, A. M., and Rosenbaum, P. R. (2000) Asymptotic separability in sensitivity analysis, {\it Journal of the Royal Statistical Society, Series B} {\bf 62}, 545--555. 

\bibitem{genz2009}
Genz, A. and Bretz, F. (2009), {\it Computation of Multivariate Normal and t Probabilities}, New York: Springer. (\texttt{R} package \texttt{mvtnorm})

\bibitem{gilbert2003}
Gilbert, P., Bosch, R., Hudgens, M. (2003). Sensitivity analysis for the assessment of the causal vaccine effects on viral load in HIV vaccine trials. {\it Biometrics} {\bf 59}, 531--541.

\bibitem{goeman2012}
Goeman, J. J., Finos, L. (2012). The inheritance procedure. {\it Statistical Applications in Genetics and Molecular Biology}, {\bf 11}, 1-18.

\bibitem{hansen2012}
Hansen, B. B. and Klopfer, S. O. (2012). Optimal full matching and related designs via network flows. {\it Journal of Computational and Graphical Statistics} {\bf 15},
609--627.

\bibitem{hayter1994}
Hayter, A. J. and Hsu, J. C. (1994). On the relationship between stepwise decision procedures and confidence sets. {\it Journal of the American Statistical Association} {\bf 89}, 128--136.

\bibitem{hosman2010}
Hosman, C. A., Hansen, B. B. and Holland, P. W. H. (2010). The sensitivity of linear regression coefficients' confidence limits to the omission of a confounder. {\it Annals of Applied Statistics} {\bf 4}, 849--870.

\bibitem{hsu2013}
Hsu, J. Y., Small, D. S., Rosenbaum, P. R. (2013). Effect modification and design sensitivity in observational studies. {\it Journal of the American Statistical
Association} {\bf 108}, 135-48.

\bibitem{hsu2015}
Hsu, J. Y., Zubizarreta, J. R., Small, D. S. and Rosenbaum, P. R. (2015). Strong control of the familywise error rate in observational studies that discover effect modification by exploratory methods. {\it Biometrika} {\bf 102}, 767--782.

\bibitem{lehmann1975}
Lehmann, E. L. (1975). {\it Nonparametrics}. San Francisco: Holden-Day.

\bibitem{lehmann2005}
Lehmann, E. L. and Romano, J. (2005). {\it Testing Statistical Hypotheses}. New York: Springer.

\bibitem{liu2013}
Liu, W., Kuramoto, J. and Stuart, E. (2013). Sensitivity analysis for unobserved confounding in nonexperimental prevention research. {\it Prevention Science} {\bf 14}, 570--580.

\bibitem{marcus1076}
Marcus, R., Peritz, E. and Gabriel, K. R. (1976). On closed testing procedures with special reference to ordered analysis of variance. {\it Biometrika} {\bf 63}, 655--660.

\bibitem{maritz1979}
Maritz, J. S. (1979). Exact robust confidence intervals for location. {\it Biometrika} {\bf 66}, 163--166.

\bibitem{obrien1987}
O'Brien, P. C. and Fleming, T. R. (1987). A paired Prentice-Wilcoxon test for censored paired data. {\it Biometrics} {\bf 43}, 169--180.

\bibitem{rosenbaum2002}
Rosenbaum, P. R. (2002). {\it Observational Studies}. New York: Springer.

\bibitem{rosenbaum2004}
Rosenbaum, P. (2004). Design sensitivity in observational studies. {\it Biometrika} {\bf 91}, 153--64.

\bibitem{rosenbaum2007}
Rosenbaum, P. R. (2007). Sensitivity analysis for m-estimates, tests and confidence intervals in matched observational studies. {\it Biometrics} {\bf 63}, 456--464. (\texttt{R} package \texttt{sensitivitymv})

\bibitem{rosenbaum2009}
Rosenbaum, P. R. and Silber, J. H. (2009). Sensitivity analysis for equivalence and difference in an observational study of neonatal intensive care units. {\it Journal of the American Statistical Association} {\bf 104}, 501--511.

\bibitem{rosenbaum2010}
Rosenbaum, P. R. (2010). {\it Design of Observational Studies}. New York: Springer.

\bibitem{rosenbaum2012}
Rosenbaum, P. R. (2012). Testing one hypothesis twice in observational studies. {\it Biometrika} {\bf 99}, 763--774.

\bibitem{rosenbaum2015}
Rosenbaum, P. R. (2015). Bahadur efficiency of sensitivity analyses in observational studies. {\it Journal of the American Statistical Association} {\bf 110}, 205--217.

\bibitem{rosenbaum2017}
Rosenbaum, P. R. (2017). {\it Observation and Experiment}.  Cambridge, MA: Harvard.

\bibitem{shaffer1986}
Shaffer, J. P. (1986). Modified sequentially rejective multiple test procedures. {\it Journal of the American Statistical Association} {\bf 81}, 826--831.

\bibitem{stuart2013}
Stuart, E. A. and Hanna, D. B. (2013). Should epidemiologists be more sensitive to design sensitivity? {\it Epidemiology} {\bf 24}, 88--89.

\bibitem{wager2017}
Wager, S. and Athey, S. (2017). Estimation and inference of heterogeneous
treatment effects using random forests.  {\it Journal of the American Statistical Association}, to appear.

\bibitem{Zhao2017}
Zhao, Q., Small, D. S. and Ertefaie, A. (2017). Selective inference for effect modification via the lasso.  {\it arXiv}:1705.08020.


\bibitem{zubizarreta2013}
Zubizarreta, J. R., Cerd\'{a}, M. and Rosenbaum, P. R. (2013). Effect of the 2010 Chilean earthquake on posttraumatic stress. {\it Epidemiology} {\bf 24}, 79--87.

\end{thebibliography}

\begin{thebibliography}{}

\bibitem{hsu2013}
Hsu, J. Y., Small, D. S., Rosenbaum, P. R. (2013). Effect modification and design sensitivity in observational studies. {\it Journal of the American Statistical
Association} {\bf 108}, 135-48.

\bibitem{hsu2015}
Hsu, J. Y., Zubizarreta, J. R., Small, D. S. and Rosenbaum, P. R. (2015). Strong control of the familywise error rate in observational studies that discover effect modification by exploratory methods. {\it Biometrika} {\bf 102}, 767--782.



\bibitem{zaykin2002}
Zaykin, D. V., Zhivotovsky, L. A., Westfall, P. H., and Weir, B. S. (2002). Trucated product method of combining $P$-values. {\it Genetic Epidemiology} {\bf 22}, 170--185.


\end{thebibliography}
\end{document}